# Optimal Placement of Mix Zones in Road Networks

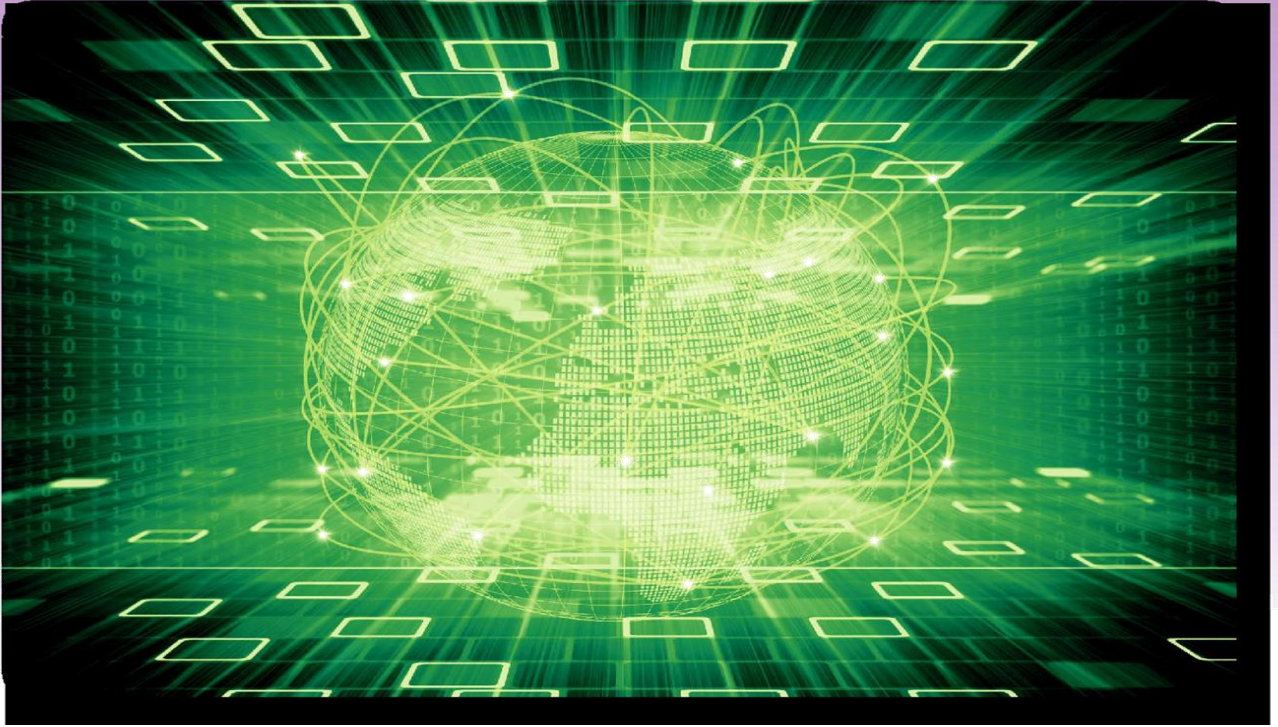

**Imran Memon**
*College of Computer Science, Zhejiang University, CHINA*
**Qasim Ali Arain**
*Beijing University of Posts and Telecommunications (BUPT), P.R, China*

*Abstract*— The road networks, vehicle users could enjoy numerous kind of services such as location based service in vehicle users can connected to Internet and communication of different users. Therefore, in order to acquire adequate privacy level and quality of service, one must have to wisely place mix zones to connect vehicle users to internet or some other internetwork. According to this research, we have analyzed the problem of optimal placement mix zones over road network. To enhance the coverage capacity of vehicles, in order to reduce the cost and communication delay. Further, it has also been discovered to minimize the cost of mix zone placement. Moreover, it has also been shown that, as the best deployment mix zones get minimized cost while at the same time the average capacity of mix zone can be maximized also privacy level increased because of optimal placement and high traffic environment.

Corresponding Author: Imran Memon (e-mail: imranmemon52@zju.edu.cn).

I. Introduction

WIRELESS networks are growing and acquired good reputation because of their less cost and easy installation.
It has been observed that, a lot of users have used wireless networks in their daily lives. As IEEE standards 802.11 maintained wireless router is very common in daily usage for internet connectivity at homes [1,2]. A lot of developments have been made in multiple wireless networks i.e. intelligent transport system which is growing very rapidly nowadays. According to road networks, certain devices are mounted on the road, which might act as access point (road side unit). Few of the RSUs are being connected to the mix zone servers which in turns have access to the internet so that people on the roads may enjoy internet service. It can also be possible to place some web cameras over the Aps to look after traffic situation, so that one can react to accident abruptly. Moreover, vehicles may possess some wireless devices which in turn communicate with other vehicles and also with Road side units.

The said network is known as road Network .It is further discovered that, agency named Federal communication Commission has given 75MHz of spectrum and dedicated short range communication among vehicles and road side units 5.9 GHz for connections [3].According to this research, we have defined the scenario, in which RSUs may cover the complete roads so that every vehicle can access RSUs without any disconnection within coverage areas. According to this construct, there is no limitation of vehicles however mobile users inside vehicles may always be allowed to access the internet.

We can define road networks as one/two direction(s) network, as we may say network along highway, or we may say multiple directional network that may cover every road in an area. In both situations, Users may be connected to internet, therefore we have to place mix zone server around the network. It seems very obvious that, to provide connectivity among RSUs and mix zones are very much important, rather then we may want to place few mix zone servers inside the network. Therefore, the problem is how to strategically place the mix zone server according to some specific requirement. A lot of problems exist in some other networks. In this paper, we have analyzed the problem of strategically placement of mix zones in both one/two and three way directional road networks. According to this idea, we deployed fix mix zones in number of user pass through near intersection and mix zones might get reduced. The main aim behind this construct is to reduce communication delay in road networks, as the road networks are related with several number of road side unit, therefore the strength of every RSU might get increased as the number of intersection among RSU and mix zones get minimized[4,5].

According to, 1-D roads, we analyzed that the strategically assignment of best deployment to reduce the number of intersection among RSUs and mix zones, while there would be only one mix zone or many mix zones accordingly. Moreover, we also have investigated the strategic assignment of
RSUs to reduce the entire cost usage. According to 2-direction road networks, we analyzed two cases, the first one is for high traffic road network, during which is decide to optimum placing of mix zones. Therefore for these larger networks, we have distributed them into minor clusters and strategically placement of one mix zone inside each cluster. So forth for the 2-directions road network, we analyzed strategic placement of multiple mix zones. According to both scenarios, our main goal is to reduce cost usage. As it is very much obvious from the previous discussion that the techniques used would to be realistic for the problems in 3-directions road networks.

Finally we analyzed the cost usage problem at placement mix zones in road network for many reasons. The first point is, if one can travel along highway in Atlanta map there might not be road side lights, hence it will be difficult to energies the RSUs. Moreover in some situation like during car racing , there is a need of temporary road networks , So in these cases privacy level of user achieve have great importance.
We have organized next parts of our paper as related work in section II. We will analyze optimal placement of mix zones in section III. During section III, we shall present the problem and investigate how to strategically placement mix zones over road network in various directions. Inside section IV we shall demonstrate problems regarding three direction road networks. Section V presents, how we should maximized placement of mix zones and minimized cost satisfy the user privacy level. Conclusion have been given in section VI

II. Related Work

Recently road network have been emerged as a latest research area. Few work done, placing of mix zone in road network is yet to be an unsolved problem in the literature. According to [6], researcher has studied, how to best placement of mix zones competently in road network. Researcher in [7,8] has given a model of mix zones in road networks to placement of mix zone due the high cost. Moreover, researcher in [9] analyzed the issue of mix zone placement optimization in participate sensing. In this work, they have investigated the optimal placement of mix zone to rectangular shape of mix zone and uniform and non-uniform traffic however at the same time practical assumption is nonrealistic and very hard maintain quality of service .Previous schemes did not consider cost factor, heavy traffic scenarios and various kind of directors mix zones.

Moreover, the problem to place mix zones strategically is very much resemble to user location privacy problem, which could be back tracked to 21 century. Researcher in 2009 has given a problem: According to that, how one can find a point, so that the distances among three vertices of a triangle get reduced. Further Torricelli and Steiner have analyzed the issue in 1644 and 1837, accordingly, and established few geometric styles. This might be optimal point might be known as Fermat's Point.

Further, this issue has been extended by [10]. Therefore, the generalized issue has been articulated model to optimal location of mix zones, this might be intended to facilitate some users that can be placement of mix zone inside a city. Hence it has to minimize the function as given below

$$f(y) = \sum_{i=1}^{m} p_i d(y - y_i)$$

Where $p_i$ are positive scalars, $y_i$ are given vectors in $R^2$, and $d(x)$ is the Euclidean norm of $x$.

Researcher in [11] have analyzed the issue while they have one as well multiple facilities. The $d(\cdot)$ in equation(1) is the Euclidean distance that discover the optimal positions of proposed an algorithm .

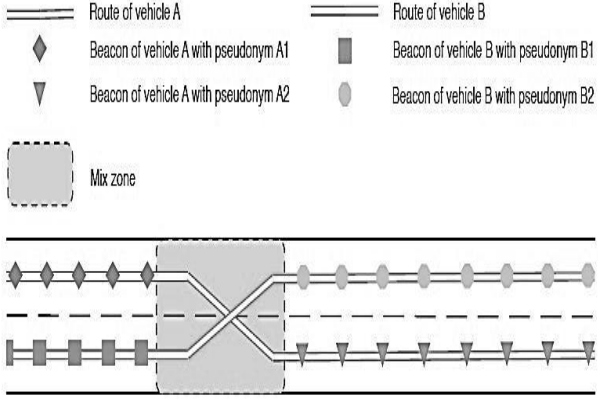

Fig. 1. Mix zone deployed over road network

It has given few analytical outcomes when there would be multiple facilities .Moreover , it has been observed that , user distances to mix zone is responsible for the distance among multiple facilities , that might not be the case during our scenario , for the consideration of placement of mix zones in road network[12].

Further, inside this manuscript, we used to reduce the cost and maximum privacy level best placement of mix zone. If we consider 1-D network, it might be little different from the facility problem, in which distance is taken into consideration, however, the number of intersection are not directly proportional to the distance. Therefore, we will explain further in section III , that it is not necessary that all link of same length to reduce the total cost. We further deduced certain analytical outcomes to place multiple mix zones in 1-D network.

Furthermore, in 2-D network scenario, we have given two novel algorithm named as optimal placement of mix zones and traffic flow mix zone placement locations accordingly, they might be used to find the optimal placement of mix zones in a high traffic road network. Therefore, if we consider larger road network we have to placement of mix zones maximization over the road network traffic passes into mix zones [13], [14], [15] and then strategically place mix zone only best location due to passed heavy traffic . We have further deduced few analytical outcomes for optimally placing multiple mix zones in 2-D network. However, reduction in total cost has also been discussed[16].

### III. Optimal Placement of Mix Zones in Road network Networks

There are many scenarios where mix zones are placed in one direction. Mix zone can deploy along a road or a highway in Fig. 1. Every mix zone is associated through neighboring intersection to certify network connectivity, and some of them are connected directly various routes in the road network. The automobiles on the road network might access the Internet by road side unit[17]. Here, we shall observe the optimum placement of mix zones so that the average number of mix zones can be placed from various routes to maximum coverage for vehicles and minimize the cost of construction in the road network.

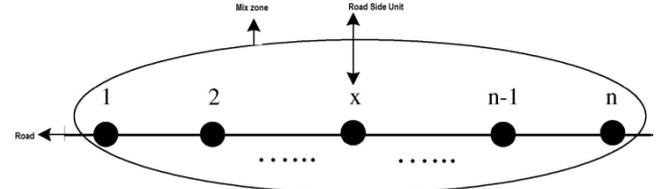

Fig. 2 one direction road network with single mix zone

#### A. one direction road network

Fig. 2,3 there are *N intersections* along a road from 1 to *N*, at least one intersection that is linked to mix zone. We denote this intersection by *I*. Intersection I is connected to the mix zone through road network . So, we consider the number of intersection connect I to the various mix zone. Let $M(I)$ be the average number of mix zones from all the intersection

$$M(I) = \frac{1}{N}[(I-1) + (I-2) + \cdots + 1 + 0 + 1 + 2 + \cdots + (N-I)]$$

$$= \frac{1}{N} \cdot \left[\left(I - \frac{N+1}{2}\right)^2 + \frac{N^2+1}{4}\right]$$

To minimize M(I), we have

$$I = \left[\frac{N+1}{2}\right] \text{ or } I = \left\lceil\frac{N+1}{2}\right\rceil \quad (2)$$

And

$$M_{min}(I) = \begin{cases} \frac{N}{4}, & N \text{ is even} \\ \left(N^2 - \frac{1}{4N}\right), & N \text{ is odd} \end{cases}$$

to connect from different route in the road network. Then, we obtain From (2,3), we located that best placement of mix zone minimum cost and passed many user through mix zone

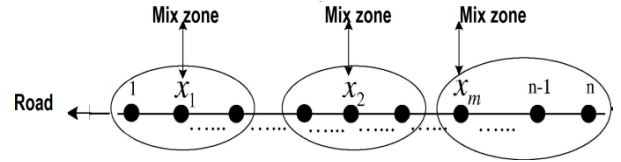

Fig. 3. one direction road network with multiple mix zones

#### B. Two directional road network

Further, we have to consider situation when there might be more than one direction over road networks. As revealed in figure. 3, there might be *N* intersection, MZ(*MZ* > 1) of which are coupled to multiple mix zones with RSU. We represent these *MZ* intersection by $I_i$, wherever $1 \le i \le MZ$ and $1 \le I_i \le N$. Every intersection always connects to mix zone and mix zones transmits to and accepts packets by the adjacent road side unit. Supposing that here are $N_i$ road side unit routing packets through the mix zones that is connected to intersection $I_i$, then

$$\sum_{i=1}^{MZ} N_i = n$$

We have

$$M(I) = \frac{1}{N}\sum_{i=1}^{MZ} MZ_i(I_i)$$

Wherever $MZ_i(I_i)$ is the number of mix zones from the $N_i$ number of roads to their mix zones coupled to intersection $I_i$, and $M(I)$ is the number of mix zones from intersection with road side unit.

1) When $C_{mv}$ = N/MZ is an odd integer, then

$$M(I) \geq \frac{1}{N}\left\{\left[\frac{N_1^2-1}{4}\right] + \left[\frac{N_2^2-1}{4}\right] + \cdots + \left[\frac{N_{MZ}^2-1}{4}\right]\right\}$$

$$\frac{1}{N}\left\{\frac{N_1^2 + N_2^2 + \cdots + N_{MZ}^2}{4} - \frac{MZ}{4}\right\}$$

$$\geq \frac{MZ}{4N} \cdot \left(\frac{N_1 + N_2 + \cdots + N_{MZ}}{MZ}\right) - \frac{MZ}{4N}$$

$$= \frac{N}{4MZ} - \frac{MZ}{4N}$$

The smallest value can be acquired when $N_i$ = N/MZ = $C_{mv}$.

2) When $C_{mv}$ = N/MZ is an even integer, then

$$M(I) \geq \frac{1}{N}\left\{\left[\frac{N_1^2}{4}\right] + \left[\frac{N_2^2}{4}\right] + \cdots + \left[\frac{N_{MZ}^2}{4}\right]\right\}$$

$$\geq \frac{1}{N}\left\{\frac{N_1^2 + N_2^2 + \cdots + N_{MZ}^2}{4}\right\}$$

$$\geq \frac{MZ}{4N} \cdot \left(\frac{N_1 + N_2 + \cdots + N_{MZ}}{MZ}\right)^2$$

$$= \frac{N}{4MZ}$$

The smallest value can be acquired when $N_i$ = N/MZ = $C_{mv}$.

3) When N/MZ is not an integer but there exists $H(1 \leq H \leq MZ)$ such that $C_{mv} = (N-H)/MZ$ is an odd integer, then

$$= \frac{N^2 - H^2}{4MZN}$$

The smallest value is attained $N_i = C_{mv} + 1$, where i ∈ B (B is a subset of {1, 2, . . . , MZ} that comprises H elements), and $N_j = C_{mv}$ where j ∈ B which is

$$M(I) \geq \frac{H(C_{mv}+1)^2 + (MZ-H)C_{mv}^2}{4N} - \frac{H}{4N}$$

We first present the mix zones placement cost construction model that we might be using during research in [13], i.e.,

$$M(I) \geq \frac{1}{N}\left\{\left[\frac{N_1^2-1}{4}\right] + \left[\frac{N_2^2-1}{4}\right] + \cdots + \left[\frac{N_{MZ-H}^2-1}{4}\right]\right.$$
$$+ \left[\frac{N_{MZ-H+1}^2}{4}\right] + \left[\frac{N_{MZ-H+2}^2}{4}\right] + \cdots$$
$$\left. + \left[\frac{N_{MZ}^2}{4}\right]\right\}$$

$$\geq \left(\frac{N_1^2 + N_2^2 + \cdots + N_{MZ}^2}{4N} - \frac{H}{4N}\right)$$

However, it seems very difficult to make $N_1 + N_2 + \cdots = N_{MZ}$ the smallest value is attained when $N_i = C_{mv}$, where i ∈ B(B is a subset of {1, 2, . . .,MZ} that contains $MZ - H$ elements), and $nN_j = C_{mv} + 1$, where j ∈ B, which is

$$M(I) \geq \frac{HC_{mv}^2 + (MZ-H)(C_{mv}+1)^2}{4N} - \frac{H}{4N}$$
$$= \frac{(N-H+MZ)(N-3H+MZ)}{4MZN}$$

4) When N/MZ is not an integer but there exists $H(1 \leq H \leq MZ)$ such that $C_{mv} = (N-H)/MZ$ is an even integer, then

$$M(I) \geq \frac{1}{N}\left\{\left[\frac{N_1^2-1}{4}\right] + \left[\frac{N_2^2-1}{4}\right] + \cdots + \left[\frac{N_H^2-1}{4}\right]\right.$$
$$\left. + \left[\frac{N_{H+1}^2}{4}\right] + \left[\frac{N_{H+2}^2}{4}\right] + \cdots + \left[\frac{N_{MZ}^2}{4}\right]\right\}$$

$$\geq \left(\frac{N_1^2 + N_2^2 + \cdots + N_{MZ}^2}{4N} - \frac{H}{4N}\right)$$

4) When MZ > N, obviously, $MZ_{min}(I) = 0$, since each mix zone might be linked to one intersection at least via road side unit.

However, the outcomes that we have acquired, we deduced that, least number of mix zones from intersection to road sideunit, i.e., $MZ_{min}(I)$, is on the order of O(N/MZ). Alternatively, $MZ_{min}(I)$, When MZ increases faster than N, privacy level increases and also cost of deployment will be decreases.

*C. minimized mix zone cost construction model*

In this section, we study the problem of mix zones placement cost construction model over the road network. Our main goal is to reduce the total placement cost of mix zones by sensibly altering the distances, heavy traffic road, and estimated time of users passing through intersection among them after we obtain the optimal position of the best mix zone placement.

$$MZP_p(d) = P_A MZ(R_L, TF, MZ_s, MZ_{Ucl}, P_{seu}, Cr)\frac{1}{d^\alpha}$$

where $MZP_p$ and $P_AMZ$ are the mix zones placement position and the placement area of mix zone, respectively, $R_L$ and $TF$ are the road lines and the road traffic, respectively, $MZ_s$ and $MZ_{Ucl}$ are the size of mix zone and the mix zone users correlation respectively, $d$ is the distance between the mix zones, $P_{seu}$ ($P_{seu} \geq 1$) is the pseudonyms changing in the mix zone that is not related to mix zone placement, $Cr$ is the mix zone coverage range, and $\alpha$ is the number of path connected to mix zone.

Assume that the road side units are connect to all vehicles in road network environment, and hence, they can successfully communication to each other and above the same threshold where $P_AMZ_i$ is placement area of mix zone for the $ith$ RSU purposes. According to (4), for a mix zone of distance $d$, $MZP_p$ mixzone placement position is

$$MZP_p = P_AMZ^{thresh} d^\alpha . Z' = d^\alpha Z$$

Where $Z$ is a constant that is determined by $R_L, TF, MZ_s, MZ_{Ucl}, P_{seu}, Cr$. Intersection that is not associated to a mix zone by $H$ Then, mix zones placement cost $PH$ for relaying nearest intersection $H$ to one mix zones through other mix zone is

$$PH = \sum_{i=1}^{j} P_AMZ_i = \sum_{i=1}^{j} d_i^\alpha Z$$

from intersection H to the mix zone, and $1 \leq H \leq N, 1 \leq j \leq N - 1$. For an intersection that is associated to a mix zones directly, As a result, the total cost placement cost of mix zone $CT$ for all the mix zones placement over the road network

*1) one direction road network:* One direction road network. As shown in Fig. 2, intersection are denoted by $1, 2, \ldots, n$, and the mix zone is connected to the intersection $I$,. Let $d_i$ be the distance between mix zone MZ and $MZ\ i+1$, where $1 \leq i \leq N-1$. Then, (5) will lead to

$$\begin{aligned} CT = &(d_1^\alpha + d_2^\alpha + \cdots + d_{I-1}^\alpha).Z \\ &+ (d_1^\alpha + d_2^\alpha + \cdots + d_{I-1}^\alpha).Z + \cdots \\ &+ d_{I-1}^\alpha.Z + d_I^\alpha.Z + \cdots \\ &+ (d_I^\alpha + d_{I+1}^\alpha + \cdots + d_{I-1}^\alpha).Z \\ &+ [(d_1^\alpha + 2d_2^\alpha + \cdots + (I-1)d_{I-1}^\alpha \\ &+ [(d_{N-1}^\alpha + 2d_{N-2}^\alpha + \cdots + (N-I)].Z. \end{aligned}$$

As a result, we obtain

$$CT \geq (N-1)\left\{\left(\prod_{i=1}^{I-1} i\, d_i^\alpha\right)\left(\prod_{j=1}^{N-I} j d_{N-j}^\alpha\right)\right\}^{\frac{1}{N-1}} Z.$$

The equality holds when $id_i^\alpha = jd_{N-j}^\alpha$ for all $i \in [1, I-1]$ and $j \in [1, N-I]$. Since

$$\sum_{i=1}^{I-1} d_i + \sum_{j=1}^{N-I} d_j = P_{seu}$$

Where $P_{seu}$ is change the pseudonyms inside of mix zone over the road of interest, then the smallest value of $CT$ can be accomplished when

$$d_i = \begin{cases} \sum_{i=1}^{I-1} i^{\frac{1}{\alpha}} \\ + \sum_{i=I}^{N-1} (N-i)^{-\frac{1}{\alpha}} \\ i^{-\frac{1}{\alpha}} d_1, \\ \quad for\ i \in [2, I-1] \\ (N-i)^{-\frac{1}{\alpha}} d_1, for\ i \in [I, N-1] \end{cases} P_{seu}, for\ i = 1$$

By analyzing this outcome, we realize that to reduce overall cost CT, the intersection that are nearer to the mix zone must have the smaller distances, that might be which is stable with our perception, since these links have additional users passed through mix zones.

*2) Two directional road network:* Further, we might consider the scenario that, there might be more than one direction as depicted in fig. 3. The inspections did represented by 1, 2, . . . , N and $m(m > 1)$ are being associated to mix zones, which might be represented {$I_j$}, wherever $i \in [1, MZ]$, $x_i \in [1, N]$, that might be present at the optimal placement positions that have already been calculated derived.

Let's consider each mix zone communicates to the nearest intersection according to the amount of users passed through mix zones. Let $B_i$ denote the set of nearest intersection relaying to the users pass to mix zone and mix zone that is connected to Intersection $I_i$. Then

$$B_i = [z_{i-1} + 1, z_i], \text{ for } 1 \leq i \geq MZ$$

Where

$$z_i = \begin{cases} 0 & for\ i = 0 \\ \dfrac{I_i + I_{i+1}}{2} & for\ i \in [1, MZ-1]\ /2 \\ N & for\ i = MZ \end{cases}$$

Let $d_i$ be the distance between MZ and $MZ_{i+1}$, where $i \in [1, N-1]$. Then, (5) results in

$$\begin{aligned} CT = &(d_1^\alpha + 2d_2^\alpha + \cdots + (I_1-1)d_{I_1-1}^\alpha \\ &+ [(d_{z_1-1}^\alpha + 2d_{z_2-2}^\alpha + \cdots + (z_1-I_1)d_{I_1}^\alpha].Z. \\ &+ [(d_{z_1+1}^\alpha + 2d_{z_2+2}^\alpha + \cdots + (I_2-1)d_{I_2-1}^\alpha \\ &+ d_{z_2-1}^\alpha + 2d_{z_2-2}^\alpha + \cdots + (z_2-I_2)d_{I_2}^\alpha].Z + \cdots \\ &+ [d_{z_{MZ-1}+1}^\alpha + 2d_{z_{MZ-1}+2}^\alpha + \cdots + (I_2-1)d_{I_{MZ-1}}^\alpha \\ &+ d_{N-1}^\alpha + 2d_{N-2}^\alpha + \cdots + (N-I_{MZ})d_{I_{MZ}}^\alpha].Z. \end{aligned}$$

So, we have

$$CT \geq (N-1)\prod_{H=1}^{MZ}\left\{\left(\prod_{i=1}^{I_{H-1}} i\, d_{z_{H-1}+i}^\alpha\right)\left(\prod_{j=1}^{z_{H-1}-I_H} j d_{z_H-j}^\alpha\right)\right\}^{\frac{1}{N-1}} Z.$$

The equality holds when $id_{z_{H-1}+i}^\alpha = jd_{z_{Pseu}-j}^\alpha$ for all $i \in [1, I_H - 1], j \in [1, z_H - I_H]$, and $H, peu \in [1, MZ]$. since

$$\sum_{H=1}^{MZ}\left(\sum_{i=1}^{I_k-1} i d_{z_{H-1}+i}^\alpha + \sum_{j=1}^{z_H-I_H} d_{z_H-j}^\alpha\right) = P_{seu}$$

When $P_{seu}$ is the mix zone changing in user inside of mix zone over the road then we have

$$\sum_{H=1}^{MZ}\left(\sum_{i=1}^{I_k-1} i^{-\frac{1}{\alpha}} + \sum_{j=1}^{Z_H-I_H}(Z_H - j)^{-\frac{1}{\alpha}}\right) d_1 =$$

where $H \in [0, MZ]$.

According to above result, we observed that minimize cost construction of mix zones placement CT, the links that are closer to the intersection connected to mix zone should have shorter distances.

### IV. OPTIMAL PLACEMENT OF Mix zones IN Three and more than three directions Road NETWORKS

In this portion, we are going to analyze the issue to find the optimal placement of mix zone in 3-D road networks, so that the number of hopes from APs to the mix zone must be reduced.

#### A. Heavy traffic Road network

It is concluded that, it would be very hard to search optimal placement of multiple mix zones inside a larger area which contain more than three directions over road network. Nevertheless, for few situations, we can first divide a larger network into multiple intersections, while determining the optimal placement of mix zones for the best place. Figure 4 is going to represent the optimal position of mix zone in a heavy traffic road in Fig. 4.

Lets assume the location of intersection by $(y_i, z_i)$, where $1 \le i \le N$. Let $MZ = (y, z)$ represent the location of the mix zone and $A$ represent the area where mix zone are located. Let's consider the mix zone is essentially linked to an intersection over road network. Then, the average number of mix zones placement position from RSU to intersection, which is denoted by Avg(MZ), satisfies

$$Avg(MZ) \ge \frac{1}{N}\sum_{i=1}^{N}\left[\frac{\sqrt{(y-y_i)^2 - (z-z_i)^2}}{r(N)}\right]$$

However $r(N)$ is the general connection series for all $N$ intersection that can guarantee connection.

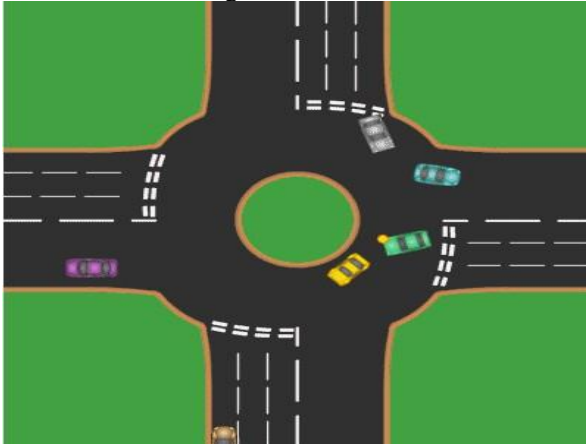

Fig. 4. Mix zone with traffic scenario

Intersection can connect to the mix zone path. However, for a larger traffic volume road network, we can deduce that the equality in (6) holds. Thus, to reduce Avg(MZ), it is equal to

Minimize

$$\left\{f(y,z) = \sum_{i=1}^{n}\sqrt{(y-y_i)^2 + (z-z_i)^2}\right\}$$

s.t $(y,z) \in A$

Enchanting the partial derivative of f(y,z) with esteem to y and z, we get

$$\frac{\partial f(y,z)}{\partial y} = \sum_{i=1}^{n}\frac{2(y-y_i)}{\sqrt{(y-y_i)^2 + (z-z_i)^2}}$$

$$\frac{\partial f(y,z)}{\partial z} = \sum_{i=1}^{n}\frac{2(z-z_i)}{\sqrt{(y-y_i)^2 + (z-z_i)^2}}$$

We getting optimal position mix zones, we want

$$\frac{\partial f(y,z)}{\partial y} = 0 \quad \frac{\partial f(y,z)}{\partial z} = 0.$$

$$A(y,z) = (0,0)$$

Where

$$A(y,z) = \left[\frac{\partial f(y,z)}{\partial y}, \frac{\partial f(y,z)}{\partial z}\right]$$

If $(y,z) \ne (y_i, z_i)$ for $i = 1, 2, \ldots, n$, and

$$A(y,z) = \begin{cases}\left(\frac{\omega k-1}{\omega k}\right) Tk, \frac{\omega k-1}{\omega k} Hk, & \omega k > 1 \\ (0,0) \end{cases}$$

or $\omega k < 1$

If $(y,z) = (y_k, z_k)$ for some $k, k [1, n]$, so

$$Tk = \sum_{\substack{i=1 \\ \ne k}}^{n}\frac{(y-y_i)}{\sqrt{(y-y_i)^2 + (z-z_i)^2}}$$

$$Hk = \sum_{\substack{i=1 \\ \ne k}}^{n}\frac{(z-z_i)}{\sqrt{(y-y_i)^2 + (z-z_i)^2}}$$

$$\omega k = (Tk^2 + Hk^2)^{1/2}$$

$$\sum_{i=1}^{n}(\vec{zz_i}/|\vec{zz_i}|) = 0$$

i.e.,

$$\sum_{i=1}^{n}\cos \alpha_i + j\sum_{i=1}^{n}\sin \alpha_i = 0$$

where $\alpha = \arg \vec{zz_i}$, i.e., the augment of $\vec{zz_i}$, and

$$\cos \alpha_i = \frac{\text{Re}(\vec{zz_i})}{|\vec{zz_i}|} \quad \sin \alpha_i = \frac{\text{Im}(\vec{zz_i})}{|\vec{zz_i}|}$$

In which Re (·) and Im (·) represent the complex numbers in terms of real and imaginary parts accordingly.

So we shall represent (11) with the help of following. Let's assume that on a plane which is horizontal named $D$, there might be an user distance which resemble to n unit force, and the value of 1, that might come from the position $z_i (1 \leq i \leq n)$. The definition of , describe that the user might get move to a location that is going to reduce its potential energy, while it will acquire its equilibrium , henceforth (11) holds.

(8)

Therefore, we find the algorithm to calculate the best locations of mix zone as described in Table-1.Moreover an illustration represented in fig 5.

We used to examine the presentation of optimal placement algorithm implemented in Matlab and also using network simulator .We generate 10000 users and also 1600 location to place 400 mix zones in the premises of 10 x 10 kilometer by using poison point process, moreover we also find the optimal location by using the algorithm. During our simulation setup, and we have organized the parameter threshold equivalent to the parameter step. Therefore, the set of steps from the start to the optimal placement mix zones can be seen in fig 6.How ever in Fig 6(a), we have to make constant the set of steps while increase number of mix zones also decreases cost. result shows that the number of mix zone place in road network and reduced cost as the value . This represent that best location optimal to placement of mix zone over . The road network then optimal placement of best location would achieved.

TABLE I
ALGORITHM FOR find best locations of mix zone

**Algorithm 1 Find best locations of mix zones**

1: **Step 1) Initialization:** we begin to search from this position
$y \sum_{i=1}^{n} \frac{y_i}{n}, Z = \sum_{i}^{n} \frac{z_i}{n}$
2: Iteration compare $(y, z)$ with $(y_i, z_i)$
3: **IF**$(y,z) \neq (y_i, z_i)$, for $i = 1,2,\ldots,n$
4: $\sum_{i=1}^{n} \cos \alpha_i + j \sum_{i=1}^{n} \sin \alpha_i = 0$
5: Find locations of mix zone randomly
6: Search best location random
7: Perform genetic operators:
   GeneticOperation($p_c$, $p_m$);
8: Perform local search:
   $P_{new}$ ! LocalSearch (locations);
9: Update new
   $P$ ! Updatelocations ($P$, $P_{new}$ );
10: Update the best individual $P_{bestlocations}$ ;
11: **Select** best locations
12: **Stopping criterion**: If $L <$ maxgen, then $L = L + 1$ and go to **Step 2**), otherwise, stop the algorithm and output.

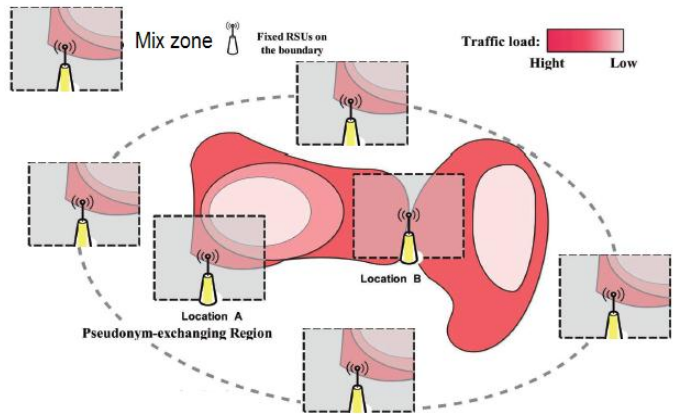

Fig. 5.find best locations of mix zone

Henceforth, Therefore, a very optimal solution can be adopted if we are going to put the mix zone near to the optimal position algorithm 2 in table. II Alternatively, we can also achieve a very sound solution by selecting the relative values in the algorithm, which can be, *step* and *best locations in Fig. 5.*

Moreover , by looking at the simulation results , it has been deduced that, when users are mix zones placed in a region by following best location passed high traffic the, then the optimal position would reach the average position as shown below.

the find best location of mixzone algorithm. However ,it can be seen that genetic algorithm is very intelligent while the results are being computed on the different time scale of 1minute to 120 minutes on a system having 2..4-GHz CPU and one 8GB RAM.

However, we have analyzed the substitute techniques that were given by 10].Moreover, we have given a new modification to the partial derivative technique that has been presented in [5]. i.e.,

$Z(y,z) = (0,0)$

$$Z(y, z) = \sum_{i=1}^{n} \frac{(y - y_i)}{\sqrt{(y - y_i)^2 + (z - z_i)^2}}$$

$$\sum_{\substack{i=1 \\ \neq k}}^{n} \frac{(z - z_i)}{\sqrt{(y - y_i)^2 + (z - z_i)^2}}$$

and ( would be a random slight positive constant. It can also be seen that (13) is permanently defined, and when ( → 0, T (x, y) → ((∂f(x, y)/∂x), (∂f(x, y)/∂y)), might be the actual description of partial derivative.

As discussed previously, we have devised an algorithm that would find the optimal placement of mix zones depends upon best location of mix zone table I, as represented in Table II.

The simulation have been performed to check the performance using genetic algorithm. We have placed the nodes in a region of 10 x 10 by following the Poisson Point process, by using 10,0000 users. So forth, the value of step is devised to be equivalent to the parameter *400 mix zones* which might be 500 RSU. The repetition times, that might be the set of steps from the start point to the optimal position, having number of mix zone (, as represented in Fig. 6(a). It is obvious from the discussion that as (reduces, tracking attacks would meet to number of mix zone increases. So forth, when number of mix zone increases attacker successful ratio decreases , would achieve high privacy level reduces attack probabilities. as represented in Fig. 6(b). we can observed privacy level user various directions, three direction road more privacy level than other because of the number of user meets mix zone over road network

*B. Normal traffic Network*

It is well known fact that, there are some examples during which 2-D road network act as normal traffic road network. As it is already been represented in in Fig 8, in which road networks are being deployed mix zone to manage many intersecting roads. Moreover, the services like VoIP, IP TV and email services might be supported. The network structure of this topology by using two mix zones has been shown in Fig.9.Let's consider, there might be set of mix zones and m of them are being connected to road side unit. For the sake of detailed analysis, we consider that all RSUs would be using the same transmission range $r(n)$,that might be equivalent to the inside the mix zones. Therefore, we have to search out the optimal positioning of those mix zones , henceforth , the number of mix zone from RSU increases must be increases privacy level.

The routing protocol selects the best route which contains lowest number of hop count among RSU $z_i$ and mix zones $g_k$, that might be represented by $H_{i,k}$.

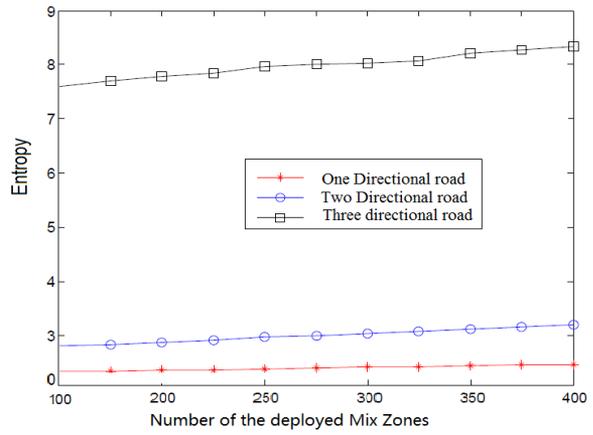

Fig. 6. (a) privacy level measured number of mix zones

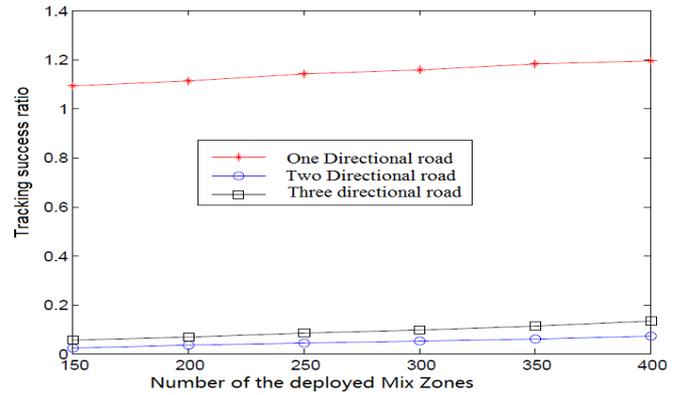

Fig. 7. Attacker tracking success

TABLE II optimal placement of mix zones

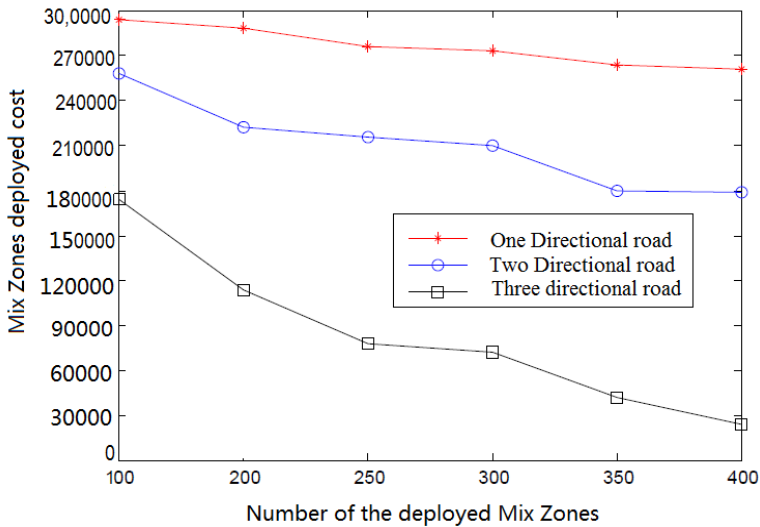

Fig. 6. (a) mix zone placement optimized deployment cost

| Algorithm 2 optimal placement of mix zones |
| --- |
| 1: **Step 1) Initialization: we begin to search from this position** $y \sum_{i=1}^{n} \frac{y_i}{n}, Z = \sum_{i}^{n} \frac{z_i}{n}$ |
| 2: see **Algorithm 1** for more information; |
| 3: **for** $i$ from 1 to mix zones placement **do** |
| 4:    **for** $j$ from 1 to $k$ **do** |
| 5:      **if** $rand^{\wedge}$ location **then** |
| 6:       select a random locations different from each other in from the *mix zone* placement *MZ* ; |
| 7:      **end if** |
| 8:    **end for** |
| 9: **end for** |
| 10: **for** $i$ from mix zones places to **do** |
| 11:    select $k$ different location from the *road network*; |
| 12: **end for** |

Henceforth, the number of mix zone that might be from RSU over road network, can be represented by MZ(g1, . . . , gm), is

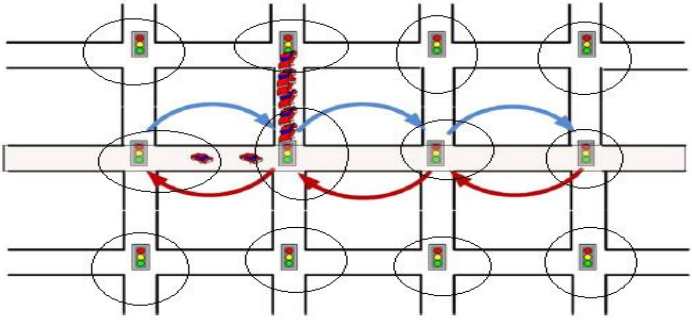

Fig. 8 optimal placement mix zones

Henceforth, to find an optimal position for m gateways inside a 2-D network might get transformed Henceforth, to find an optimal position for m mix zone inside a 2-D network might get transformed into a into a 1-D network in Fig. 8 and 9, which might be resolved using the technique that is very much analogous to those discussed in Sections III and IV.

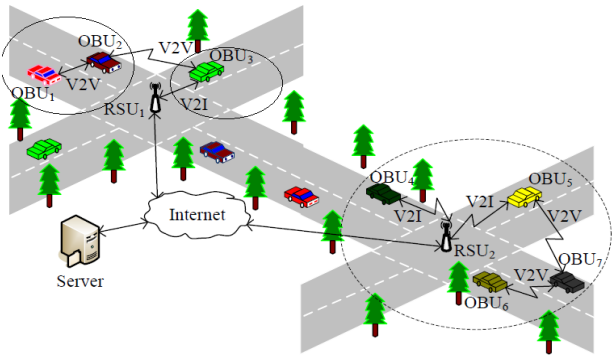

Fig. 9 road network scenario

. *Discussions*

It has been shown in Figure .7 ·. The *tracking success* is the percentage of MOs that can be tracked over $j$ consecutive Mix-Zones. If $M_s$ is the number of MOs successfully tracked and $M$ is the total number of MOs that have traversed $j$ consecutive Mix-Zones, then, the tracking success $TS(j) = M_s(j)/M(j)$ measured as percentage as shown in Figure .7 ·

The *cumulative entropy* of a particular MO $m$ on the other hand is

$$H(m, J) = \sum_{i=1}^{J} H_i(m) \times m.$$

Here, $J$ is the total number of Mix-Zones traversed by MO $m$ over the said road network area.

According, to these both networks as represented in IV-A and B, accordingly. Therefore, the mix zone cost *CT* for placement of mix zones among road network and depends on user passes through mix zones can be shown below

$$CT = pL \cdot I(G) \propto r^\gamma(n) \cdot I(G)$$

Wherever $P_L$ is going to represent the location placement, that might be equivalent number of RSUs coverage, where G represent the counter of mix zones inside the network, $I(G)$ is going to show the number of intersection among an RSU and a mix zone, and γ can be the path loss exponent. Therefore, as the I(G) get near coverage and high privacy then the Mix zone cost CT might also get reduced. However, if the cost get directly proportional to the distance, then it might also be directly proportional to the number of place mix zone, henceforth it has been assumed that every mix zone has the various length. Therefore, if best location mix zone placed so reduced cost, then the overall cost would also be reduced. However, the techniques that might be represented in this section might also be applied to 3-D space.

V. MORE DISCUSSIONS

In the last discussion, we have analyzed the issue of optimal placement of mix zones in road network, so forth, the number of RSU connect from intersection to mix zones have been reduced. However, there is other lot of advantages of our research.

Moreover

Moreover, a very important factor that is known as end to end delay have been reduced and increases user privacy over the road network . As it is obvious from the previous discussion that communication delay in wireless networks because of disputation, however, to process and query during every intermediate user and also the number of RSU counts are the better metrics for it .Therefore, if the number of mix zone placed best locations increases privacy level and communication delay as well as cost are being reduced, resulting overall cost would also be reduced.

Moreover, it is also been observed that, the average road side unit might get increased. Let's consider that every RSU would send and receive data traffic at the same rate, which might be represented by $\varphi$. We may represent the total intersection strength of the road network by $I_T$ that would be the same as stated by [11]. Therefore, we have

$$\varphi \cdot I(x) \cdot n \leq I_T \quad (14)$$

Wherever $I(x)$ is going to represent the average number of intersections count, and $n$ might be the total number of RSUs in the network. From (14), we might attain

$$\varphi \leq \frac{I_T}{n\,I(x)} \leq \frac{I_T}{n I_{min}(x)}$$

Wherever $I_{min}(x)$ can be the least average number of intersection count, as already derived. However, it can be easily analyzed that, as the average number of RSU count get reduced, the strength of mix zones might get increased.

VI. CONCLUSION

According to this research, we have resolved the issue of optimal placement of mix zone more than one directions ( inside 1 ,2D and multiple directions) over road networks to reduce the number of mix zone through best placement of mix zone and reduce the deployment cost .Therefore, the delay in communication might get reduced and increases privacy level .However the strength of each and every mix zone might get increased. So forth, in 1-D roads, we have been received the investigative outcomes for the placement of optimal mix zones. Moreover, in 2-D and multiple direction roads, we have designed two algorithms, they might be known as the best location of mix zone and optimal placement of mix zones accordingly, in order to obtain the optimal placement of the mix zone in road networks.

According to our simulation results, it has been revealed that our technique has outperformed all other techniques and shows better results.